\documentclass[preprint, 1p]{elsarticle}

\usepackage{caption}
\usepackage{graphicx}
\usepackage{subcaption}
\usepackage{placeins}
\usepackage{afterpage}

\usepackage[utf8]{inputenc}
\usepackage{amssymb,amsmath}
\usepackage{lipsum}
\usepackage{sectsty}
\usepackage{fullpage}
\usepackage{microtype}
\usepackage{acronym}
\usepackage{multirow,algorithm}
\usepackage{booktabs}

\graphicspath{{../figs/}}

\usepackage{lineno}

\usepackage{xargs} 
\newcommandx{\myparagraph}[1]{\paragraph{#1}}
\newcommandx{\mysubparagraph}{\bigskip \indent}

\usepackage[table,pdftex,dvipsnames]{xcolor}
\usepackage[colorlinks]{hyperref}
\AtBeginDocument{%
  \hypersetup{%
    linkcolor=[RGB]{0,150,0},%
    citecolor=blue,%
  }%
}%

\newcommand{\real}{\ensuremath{\mathbb{R}}}

\DeclareMathOperator*{\minimize}{\mathrm{minimize}}
\DeclareMathOperator*{\maximize}{\mathrm{maximize}}

\def \X {\mathcal{X}}

\DeclareMathOperator{\V}{\mathbf{V}}
\DeclareMathOperator{\U}{\mathbf{U}}

\DeclareMathOperator{\tenX}{\boldsymbol{\mathcal{X}}}
\DeclareMathOperator{\tenD}{\boldsymbol{\mathcal{D}}}

\DeclareMathOperator{\tenA}{\boldsymbol{\mathcal{A}}}
\DeclareMathOperator{\tenB}{\boldsymbol{\mathcal{B}}}
\DeclareMathOperator{\Pmat}{\mathbf{P}}

\def \v {{\bf v}}
\def \w {{\bf w}}
\def \u {{\bf u}}

\newcommand {\btimes} {{\times}}

\journal{Neuroimage}
\biboptions{authoryear,square}

\begin{document}
\begin{frontmatter}

\title{Tensor network factorizations:  Relationships between brain structural connectomes and traits}

			\author[RC]{Zhengwu Zhang \corref{fmr}}
			\author[Rice,BCM]{Genevera I. Allen}
			\author[MDander]{Hongtu Zhu}
		        \author[Duke]{David Dunson}

			\ead{zhengwu\_zhang@urmc.rochester.edu}

			\address[RC]{Department of Biostatistics and Computational Biology, Rochester, NY, USA}
			
			\address[Rice]{Departments of Statistics, Computer Science,
  Electrical and Computer Engineering, Rice University, 
Houston, TX, USA }

                        \address[BCM]{Neurological Research Institute, Baylor College of Medicine, Houston, TX, USA}
			
			\address[MDander]{Department of Biostatistics,
				The University of Texas MD Anderson Cancer Center,
				Houston, TX, USA}
				
			\address[Duke]{Department of Statistical Science, Duke University, Durham,NC,  USA}

\cortext[fmr]{Department of Biostatistics and Computational Biology, Rochester, NY, USA}

\begin{abstract}

Advanced brain imaging techniques make it possible to measure individuals' structural connectomes in large cohort studies
non-invasively. The structural connectome is initially shaped by genetics and subsequently refined by the environment. It is extremely interesting to study relationships between structural connectomes and environment factors or human traits, such as substance use and cognition. Due to limitations in structural connectome recovery, previous studies largely focus on functional connectomes. Questions remain about how well structural connectomes can explain variance in different human traits. Using a state-of-the-art structural
connectome processing pipeline and a novel dimensionality reduction technique applied to data from the Human Connectome Project (HCP), we show strong relationships between structural connectomes and various human traits.  Our dimensionality reduction approach uses a tensor
characterization of the connectome and relies on a generalization of principal components analysis.  We analyze over $1100$ scans for $1076$ subjects from the HCP and the Sherbrooke test-retest data set, as well as $175$ human traits that measure domains including
cognition, substance use, motor, sensory and emotion.  We find that structural connectomes are associated with many traits. Specifically, fluid intelligence, language comprehension, and motor skills are associated with increased cortical-cortical brain structural connectivity, while the use of alcohol, tobacco, and marijuana are associated with decreased cortical-cortical connectivity.   

\end{abstract}


\end{frontmatter}


\section{Introduction}

The human brain structural {\em connectome},  defined here as the collection of white matter fiber tracts connecting different regions of the brain \citep{park2013structural,fornito2013graph,craddock2013imaging,Jones2013}, plays a crucial role in how the brain responds to everyday tasks and life's challenges.  There has been a huge interest in studying connectomes and understanding how they vary for individuals in different groups according to traits and substance exposures. Such studies have typically focused on functional connectomes instead of structural connectomes \citep{park2013structural,finn2015functional,bjork2014effects,smith2015positive,price2012review}, due to the difficulty of recovering reliable structural connectomes \citep{maier2016tractography,reveley2015superficial}. Recent advances in noninvasive brain imaging and preprocessing have produced huge brain imaging datasets (e.g., the Human Connectome Project \citep{van2013wu} and the UK Biobank \citep{miller2016multimodal}) along with sophisticated tools  to routinely extract brain structural connectomes for different individuals. Relying on high quality imaging data and many different traits for a large number of study participants obtained in the Human Connectome Project (HCP) \citep{glasser2013minimal,glasser2016human}, this article focuses on analyzing relationships between brain structural connectomes and different traits and substance exposures using novel data science tools applied to the HCP data.

Estimation of the structural connectome relies on a combination of diffusion magnetic resonance imaging (dMRI) and structural MRI (sMRI).  dMRI collects information on the directional diffusion of water molecules in the brain \citep{behrens2003non, parker2003,jbabdi2015measuring}.  As diffusion tends to occur in a directional manner along fiber tracts in white matter, while being essentially non-directional in gray matter, dMRI provides information on locations of fiber tracts. Tractography \citep{Basser2000,Descoteaux2009,Girard2014266} extracts the tracts from dMRI data, yielding a very large number of `tubes' snaking through the brain.  These data are enormous and complex, and  statistical analysis is not straight forward. The difficulty comes from several aspects. First, the dimensionality of each subject's data ($p$) is extremely large (each subject usually contains $> 1$ million 3D curves), but the sample size ($n$) is relatively small. This problem is referred as the large $p$ small $n$ problem in statistics \citep{dunson2018statistics}.  Second, the data are geometrically structured and it is unclear where the signal is, i.e., the component of variation of the data that can explain trait difference in different individuals. Last but not least, alignment and building correspondence between tracts are very hard. For these reasons, it is typical to parcellate the brain into anatomical regions of interest (ROIs) using sMRI \citep{Desikan2006968,Destrieux2010} according to pre-defined templates \citep{Desikan2006968,Destrieux2010,glasser2016multi}.  This allows coarse alignment of different individuals (by mapping the template to individual space), and connectome data reduction into a connectivity matrix.

 Although reconstruction errors inevitably occur \citep{maier2016tractography,reveley2015superficial,thomas2014anatomical}, advances in imaging techniques \citep{setsompop2012improving} and preprocessing pipelines \citep{Girard2014266, Smith2012, Zhang2017HCP, donahue2016using} have improved the reconstruction of structural connectomes \citep{donahue2016using}. In this paper, we will explore whether existing structural connectome reconstruction together with proposed novel analysis methods are sufficient to detect (potentially subtle) associations between connectomes and various human traits. We will also compare the structural connectomes with other type of connectomes, eg.,  local structural connectomes \citep{powell2018local,yeh2016connectometry}, in predicting traits. This comprehensive connectome-trait analysis will give us a systematic understanding of which and how human traits are related to the structural connectome. 
  

In our analysis framework, we rely on a new structural connectome processing pipeline, and improved methods for representing brain connectomes \citep{Zhang2017HCP}.  Previous statistical approaches focus primarily on reducing the connectome to a binary adjacency matrix containing 0-1 indicators of any fiber connections between ROIs.   The adjacency matrix is then further reduced to topological summary statistics of the brain graph, providing low-dimensional numerical summaries to be used in statistical analyses \citep{watts1998cds, Durante2014n, sporns2004small}.  Such connectome simplification is appealing due to its interpretability, but leads to an enormous loss of information.  Instead, we use a {\em tensor network} representation that incorporates multiple features 
measuring the strength and nature of white matter tracts between each pair of brain ROIs.  This representation better preserves information in the tractography data, and allows flexibility in examining associations with traits.  We extend principal components analysis (PCA) to tensor network data via a semi-symmetric tensor decomposition method, which produces brain connectome PC scores for each subject.  These scores can be used for visualization and efficient inference on relationships between connectomes and human traits. The main contributions of this paper can be summarized as follows:
\begin{enumerate}
\item In conjunction with a recent dMRI preprocessing pipeline PSC (Population-based Structural Connectome  \citep{Zhang2017HCP} analysis pipeline, available at \url{https://github.com/zhengwu/PSC_Pipeline}
), this paper provides additional flexible statistical toolboxes to the neuroscience community to statistically analyze structural connectomes in large cohort studies. The toolboxes will be publicly available along with the PSC GitHub repository, providing functions of: 1) mapping high-dimensional connectomes to low-dimensional space for visualization; 2) hypothesis testing of connectome distribution difference; and 3) relating brain connectomes with human traits.

\item Based on our analysis of data on $1076$ individuals and $175$ traits, we find a strong relationship between structural cortical-cortical connectomes and multiple traits, particularly those related to cognition, motion and substance usage. Extensive predictive analyses also confirm that the structural connectomes can significantly improve the prediction of these traits in addition to demographic predictors such as age and gender.  Further investigation shows that traits related to positive lifestyles, such as 
 good reading ability, high fluid intelligence, and good motor skills, tend to have positive correlations on cortical-cortical brain connections. On the other hand, substance use, including binge drinking,  tobacco, and marijuana use, can reduce interconnections.  These findings are consistent with related analyses for functional connectomes \citep{smith2015positive,finn2015functional}.

 \end{enumerate}

\section{Methods}

\subsection{Data sets}

We focus on three data sets in this paper, which contain about $1,221$ dMRI scans for $1,067$ subjects. 

\noindent{\bf Human Connectome Project (HCP) data set}: The HCP aims to characterize human brain connectivity in about $1,200$ healthy adults and to enable detailed comparisons between brain circuits, behavior and genetics at the level of individual subjects. Customized scanners were used to produce high-quality and consistent data to measure brain connectivity. The latest release in 2017, containing various traits, structural MRI (sMRI) and diffusion MRI (dMRI)  data for $1065$ healthy adults, can be easily accessed through ConnectomeDB. The rich trait data, high-resolution dMRI and sMRI make it an ideal data set for studying relationships between connectomes and human traits.

A full dMRI session in HCP includes 6 runs (each approximately 10 minutes),
representing 3 different gradient tables, with each table acquired once with right-to-left and left-to-right phase encoding polarities, respectively. Each gradient table includes approximately $90$
diffusion weighting directions plus 6 $b_0$ acquisitions interspersed throughout each
run.  Within each run, there are  three shells of $b=1000, 2000$, and $3000$ s/mm$^2$ interspersed
with an approximately equal number of acquisitions on each shell.   The directions were optimized so that
every subset of the first $N$ directions is also isotropic. The scan was done by using the Spin-echo EPI sequence on a 3 Tesla customized Connectome Scanner. See \cite{VanEssen20122222} for more details about the data acquisition of HPC. Such settings give the final acquired image with isotropic voxel size of $1.25$ mm, and $270$ diffusion weighted scans distributed equally over $3$ shells.


\noindent{\bf  HCP Test-Retest data set}: A subset of HCP participants were recruited to undergo the full 3T HCP imaging and behavioral protocol for a second time. We identified and successfully processed structural connectome data from 44 subjects (88 scans). 

\noindent{\bf  Sherbrooke Test-Retest data set}: Different from the high-resolution HCP data set, this data set represents a clinical-like acquisition using a $1.5$ Tesla SIEMENS Magnetom. There are $11$ subjects with 3 acquisitions for each subject. A total of 33 acquisitions, from 11 healthy participants, were included. The  diffusion space (q-space) was acquired along 64 uniformly distributed directions, using a b-value of $b = 1000$ s$/$mm$^2$ and a single $b_0$ (=0 s/mm$^2$) image. The dMRI has a 2 mm isotropic resolution.  An anatomical T1-weighted $1\times1\times1$ mm$^3$ MPRAGE (TR/TE $6.57/2.52$ ms) image was also acquired. 

\subsection{Brain Connectome Extraction} 
\begin{figure}
\begin{center}
\begin{tabular}{c}
\includegraphics[height=2.0 in]{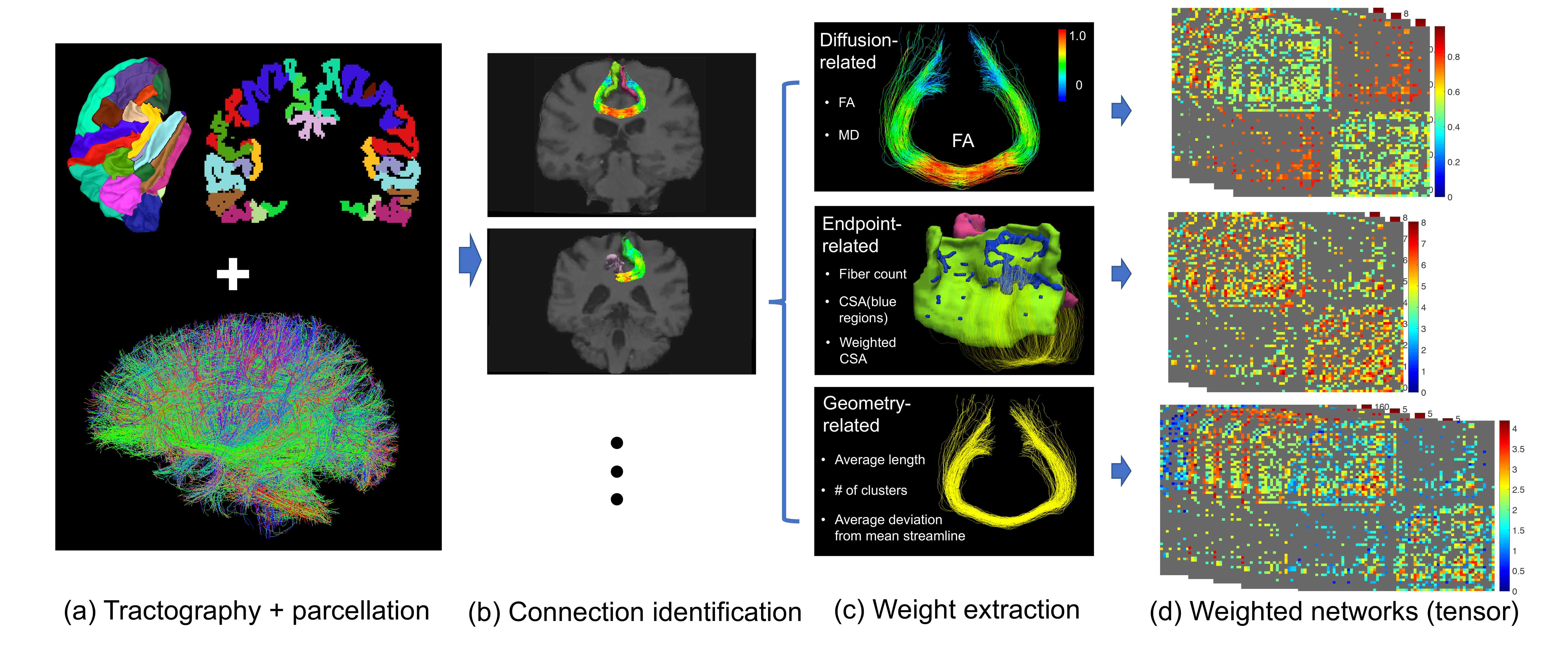} 
\end{tabular}
\caption{Pipeline of the preprocessing steps to extract weighted networks from the dMRI and sMRI image data. (a) Desikan-Killiany parcellation and the tractography data for an individual's brain; (b) extraction of streamlines between two ROIs; (c) feature extraction from each connection and (d) extracted weighted networks. } 
\label{fig:illustration}
\end{center}
\end{figure}

From the raw diffusion MRI (dMRI) and structural MRI (sMRI) data, we want to reliably extract the brain structural connectome data.  Figure \ref{fig:illustration} illustrates our data processing pipeline. 

\noindent{\bf HARDI tractography construction}:
A reproducible probabilistic tractography algorithm
\citep{Girard2014266,maier2016tractography} is used to generate the whole-brain tractography data set of  each subject  for all data sets. The method borrows anatomical information from high-resolution T1-weighted imaging to reduce bias in reconstruction of tractography. Also the parameters are selected based on evaluation of various global connectivity metrics in \citep{Girard2014266}. In the generated tractography data, each streamline has a step size of 0.2 mm. On average, $10^5$ voxels were identified as the seeding region (white matter and gray matter interface region) for each individual in the HCP data set (with isotropic voxel size of 1.25 mm).  For each seeding voxel, we initialized $16$ streamlines to generate about $10^6$ streamlines for each subject. We have similar settings for the Sherbrooke test retest data set.  

\noindent{\bf Network node definition}: We use the popular Desikan-Killiany atlas to define the brain regions of interest (ROIs) corresponding to the nodes in the structural connectivity network. The Desikan-Killiany parcellation has 68 cortical surface regions with 34 nodes in each hemisphere. Freesurfer software is used to perform brain registration and parcellation. Figure \ref{fig:illustration} column (a) illustrates the Desikan-Killiany parcellation and a reconstructed tractography data after subsampling.

\noindent{\bf Connectome tensor extraction}:  With the parcellation of an individual brain, we extract a set of weighted matrices to represent the brain's structural connectome. To achieve this goal, for any two ROIs, one needs to first extract the streamlines connecting them. Alignment of the parcellation (on T1 image) and tractography data (dMRI image) is done using Advanced Normalization Tools (ANTs) \citep{Avants20112033}. To extract streamlines connecting ROI pairs, several procedures are used to increase the reproducibility: (1) each gray matter ROI is dilated to include a small portion of white matter region, (2) streamlines connecting multiple ROIs are cut into pieces so that we can extract the correct and complete pathway and (3) apparent outlier streamlines are removed. Extensive experiments have illustrated that these procedures can significantly improve the reproducibility of the extracted connectomes, and readers can refer to  \cite{Zhang2017HCP} for more details.

To analyze the brain as a network, a scalar number is usually extracted to summarize each connection. For example, in the current literature \citep{fornito2013graph,Smith2013,Jones2013}, count is considered as a measure of the coupling strength between ROI pairs. However, fiber count can be unreliable due to tractrography algorithms and noise in the dMRI data  \citep{maier2016tractography,fornito2013graph}.  Instead of only using the count as the ``connection strength", we include multiple features of a connection to generate a tensor network for each brain. The tensor network has a dimension of $P \times P \times M$, where $M$ represents the number of features and $P$ represents the number of ROIs.  Each of the $M$ matrices is a weighted network and describes one aspect of the connection.  As illustrated in the third column of Figure \ref{fig:illustration}, the following features are extracted. 

\begin{enumerate}
\item {\bf Endpoint-related features}: We consider the features generated from  the end points of streamlines for each ROI pair. The first feature we extract is the number of endpoints, which is the same as the count of streamlines. Another feature we extracted is connected surface area (CSA), which is proposed in \citep{Zhang2017HCP}. To extract CSA, at each intersection between an ROI and a streamline, a small circle is drawn, and the total area covered by these circles is the CSA.  A weighted version of CSA is calculated by dividing the total surface area of the two ROIs. 

\item{\bf Diffusion-related features}: Diffusion metrics,  such as FA, characterize the water diffusivity at a particular location or voxel. We extract these diffusion metrics along white matter  streamlines for each ROI pair.

\item {\bf Geometry-related features}:  Geometric features such as the average length and  cluster configuration \citep{Zhang2017HCP}
that can characterize  the   geometry of  streamlines are extracted. 
\end{enumerate}

Applying the pipeline in Figure \ref{fig:illustration}, we processed the Sherbrooke test-retest data set  with $11$ subjects and $3$ repeated scans per subject (for our reproducibility study), and the HCP data set with $1065$ subjects (from the latest release of HCP data set in 2017) \citep{van2013wu}. For each subject, $12$ weighted networks are extracted: three endpoint-related features (count of streamlines, connected surface area (CSA) and weighted CSA), four diffusion-related features (mean and max values of fractional anisotropy (FA) and mean diffusivity (MD)), and five geometry-related features (cluster number, average length and mean deviations from a template streamline).   In addition, $175$ trait measures for each subject in the HCP data set were also obtained. The $175$ trait measures are from eight categories describing a human's status and behavior: cognition, motor, substance use, psychiatric and life function, sense, emotion, personality, and health.  A detailed description of these traits is included in the Excel spreadsheet of Supplementary Material II. In total, we extract $13,176$ networks and $186,375$ trait scores for our statistical analyses.

\subsection {Tensor-Network Principal Components Analysis}

Our focus is on inferring relationships between brain structural connectomes and human traits.  To address this goal using the extracted connectome representation, it is necessary to develop a statistical approach to assess associations between tensor network representations of the brain connectome and traits.  A key problem in this respect is how to estimate low-dimensional features summarizing the brain tensor network without loosing valuable information.  If the connectome data for each individual could be reorganized into a vector, Principal Components Analysis (PCA) could be used to extract brain connectome PC scores for each individual.  However, vectorizing the network would discard valuable information about the network structure.  Instead, we propose to use a semi-symmetric tensor generalization of PCA, named tensor-network PCA (TN-PCA).

To begin, we need
some notation which is largely adapted from \cite{Kolda:2009}. 
Let ${\tenX} \in \real
^{I_1 \times I_2 \times ... \times I_N}$ be an {\it N-mode}
tensor, and we denote matrices as ${\bf X}$, vectors as ${\bf x}$ and
scalars as $x$. The 
{\it outer product} is denoted by $\circ$: ${\bf x} \circ {\bf y} =
\bf {x y^T}$. The {\it scalar product} of two tensors ${\tenA},{\tenB}
\in \real ^{I_1 \times I_2 \times ... \times I_N}$ is defined as
$\left<\tenA, \tenB \right> = \sum_{i_1}\sum_{i_2}...\sum_{i_N}
a_{i_1i_2...i_N}b_{i_1i_2...i_N}$. The {\it Frobenius norm} of a
tensor $\tenX$ is $\| \tenX \|_F = \sqrt{\left<\tenX,\tenX
  \right>}$. The {\it n-mode multiplication} of tensor $\tenX  \in
\real ^{I_1 \times I_2 \times ... \times I_N}$ with  a matrix ${\bf A}
\in \real^{J_n \times I_n}$, denoted by $\tenX \times_n {\bf A} $,
gives a tensor in $\real^{I_1 \times... I_{n-1} \times J_n \times
  I_{n+1}...\times I_N}$, where each element is the product of
mode-$n$ fiber of $\X$ multiplied by ${\bf A}$.  When ${\bf A}$
degenerates to a vector ${\bf a}$, i.e. ${\bf a} \in \real^{I_n}$, we
have $\tenX  \times_n {\bf a} \in \real^{I_1 \times... I_{n-1} \times
  I_{n+1}...\times I_N}$, which is an $N-1$-mode tensor.  
  
  For
simplicity, we will only consider three-mode tensors here, but all of
the methods can be easily extend to higher-order tensors. 
We will work with a tensor network, ${\tenX} \in \real ^{P \times P
  \times N}$, which is a concatenation of network
adjacency matrices ${\bf {A}}_i \in \real^{ P \times P}$ for
$i=1,...,N$, where $P$ represents the number of nodes and $N$
represents the number of subjects. For example, if we only use the count feature, by concatenating all subjects' count adjacency matrices in the HCP data, we get a tensor of $68 \times 68 \times 1065$    (if all features are included, we will have a tensor of  $68 \times 68 \times 12 \times 1065$).  We say that our tensor network is {\it semi-symmetric}
as every frontal
slice, $\X_{:,:,n}$, is a symmetric matrix: $\tenX_{i,j,n} =
\tenX_{j,i,n} \forall \ \ i,j,n.$

Given the special structure of our semi-symmetric tensor network,
existing tensor decompositions are not ideal for conducting TN-PCA.  Consider an extension of the popular {\it Tucker} model
\citep{tucker_1966} 
which forces the first two Tucker factors to be equal to account for
the semi-symmetric structure: \begin{align}
\label{tucker_model}
\tenX \approx \tenD \times_{1} \V \times _{2} \V \times_{3} \U,
\end{align}
where $\V_{P \times K_V}$ and $\U_{N \times K_U}$ are orthogonal matrices
that form the Tucker factors and $\tenD_{K_V \times K_V \times K_U}$ is the
Tucker core; under certain restrictions on the Tucker core, this model
will result in a semi-symmetric tensor.  Interestingly, when
standard algorithms for estimating Tucker models 
(e.g. Higher-Order SVD and Higher-Order Orthogonal
Iteration, HOSVD and HOOI, respectively
\citep{tucker_1966,de_lathauwer_2000,kolda_tensor_review}) are applied
to semi-symmetric tensors, they 
result in tensor factorizations that follow model
\eqref{tucker_model}; this fact can be easily verified and is also
discussed in \cite{de_lathauwer_2000}. Despite the ease of
implementation of Tucker 
models for semi-symmetric tensors, these approaches are not ideal for studying and embedding 
brain networks.  The semi-symmetric Tucker core makes it difficult
to directly interpret the effects and prevalence of tensor network
components (eigenvectors associated with $\V$) across the population.
Further, the assumption that the population factors, $\U$, are
orthogonal is likely overly restrictive and limits the Tucker model's
ability to fit brain connectome data well.

Because of this, we consider  another popular tensor decomposition
model: the CP decomposition, which models a tensor as a sum of rank one tensors: $\tenX \approx
\sum_{k=1}^{K} d_{k} \v_{k} \circ \v_{k} \circ \u_{k}$
\citep{carroll_cp_1970,harshman_1970}.  As with the Tucker model, it is clear that the first two factors
must be equivalent to yield a CP model appropriate for semi-symmetric
tensors:
\begin{align}
\label{cp_model}
\tenX \approx \sum_{k=1}^{K} d_{k} \v_{k} \circ \v_{k} \circ \u_{k}.
\end{align}
Here, $\v_{k}$ and $\u_{k}$ are $P$ and $N$ vectors, respectively,
that form the $k^{th}$ CP factor and $d_{k}$ is the $k^{th}$ positive
CP scaling parameter.  It is clear that \eqref{cp_model} needs
no further restrictions to yield a semi-symmetric tensor.  Yet, this
model may not be ideally suited to modeling our population of brain
connectomes.  If there are no restrictions on the CP factors $\V$ as
is typical in CP 
models, then the columns of $\V$ could be highly correlated and fail
to span the eigen-space of the series of brain networks.
Hence, we propose to add an additional orthogonality constraint on the
CP factors $\V$, but leave $\U$ unconstrained.  Note that this form of
orthogonality in one factor but not the other is distinct from the
various forms of orthogonal tensor decompositions proposed in the
literature \citep{kolda2001orthogonal}.

We estimate our CP model for semi-symmetric tensors by solving the
following least squares problem:
\begin{align}
\label{cp_opt_full}
\minimize_{d_k,\v_{k}, \u_{k}}  & \| \tenX - \sum_{k=1}^{K} d_{k} \v_{k} \circ \v_{k} \circ \u_{k}\|_{2}^{2}   \\ 
\textrm{subject to }  & \u_{k}^{T} \u_k = 1,  \v_{k}^{T} \v_{k} = 1, \v_{k}^{T} \v_{j} = 0 \ \forall  j<k.  \nonumber
\end{align}
As with the typical CP problem, this is non-convex but is instead bi-convex in $\v$ and $\u$.  The most common optimization
strategy employed is block coordinate descent which alternates solving a least squares problem for all the $K$ factors, for $\V$ with $\U$ fixed
and for $\U$ with  $\V$ fixed \citep{kolda_tensor_review}.  For our problem with the additional orthogonality constraints, this approach is computationally prohibitive.  Instead, we propose to use a greedy one-at-a-time strategy that sequentially
solves a rank-one problem, a strategy sometimes called the tensor
power method \citep{de2000best,allen2012sparse}. The single-factor CP
method can be formulated as 
\begin{align}
\label{cp_opt_ours}
\maximize_{\u_k, \v_k} \ \ & \tenX \times_{1} ( \Pmat_{k-1} \v_{k}) \times_{2}
( \Pmat_{k-1} \v_{k}) \times_{3} \u_{k} \\
\textrm{subject to} \ \ & \u_k^{T} \u_{k} = 1,  \ \v_{k}^{T}
\v_{k} = 1,  \nonumber
\end{align}
where $\Pmat_{k-1} = \mathbf{I} - \V_{k-1} \V_{k-1}^{T}$ is the
projection matrix with $\V_{k-1} = [ \v_{1}, \ldots \v_{k-1}]$ denoting
the previously estimated factors.  \eqref{cp_opt_ours} uses a
Gram-Schmidt scheme to impose orthogonality on $\v_k$ via the
projection matrix $\Pmat_{k-1}$; it is easy to verify that
\eqref{cp_opt_ours} is equivalent to \eqref{cp_opt_full}
\citep{allen2012sparse}.

To solve \eqref{cp_opt_ours}, we employ a block coordinate descent
scheme by iteratively optimizing with respect to $\u$ and then
$\v$; each coordinate-wise update has an analytical solution:
\begin{align}
\label{cp_update_u} 
\u_{k} &= \frac{\tenX \times_{1} \v_{k} \times_{2}
  \v_{k}}{||\tenX \times_{1} \v_{k} \times_{2}
  \v_{k}||_{2}}, \\
\label{cp_update_v}   
\v_{k} &= {E_{max} \left( \Pmat_{k-1} ( \tenX
  \times_{3} \u_{k} ) \Pmat_{k-1}  \right)}.
\end{align}
Here, $E_{max}(\mathbf{A})$ refers to the eigenvector corresponding to the maximum eigenvalue of
matrix $\mathbf{A}$.  We can show that this scheme converges to a
local optimum of \eqref{cp_opt_ours}.
Putting together these pieces, we present our tensor power
algorithm for solving \eqref{cp_opt_full} in Algorithm~\ref{algo:tensorcp}. After we greedily estimate a rank-one factor, we use subtraction
deflation. Overall, this algorithm scales well computationally compared 
to the Tucker model which requires computing multiple SVDs of
potentially large matricized tensors.
\begin{algorithm}
\caption{Tensor power method for semi-symmetric CP decomposition
  (TN-PCA Algorithm)}
\label{algo:tensorcp}
Let $\tenX$ be a three-way tensor concatenating $M$ brain networks. The tensor power method for semi-symmetric CP decomposition of $\tenX$ is given as:
\begin{enumerate}
\item Let $\hat {\tenX} = \tenX$. 
\item For $k = 1,...,K$, do 
\begin{enumerate}
\item Find the semi-symmetric  single-factor CP decomposition for $\hat{\tenX}$. Initialize $\v_k, \u_k$, and iteratively update  $\v_k, \u_k$ until convergence:
\begin{enumerate}
\item $\u_k = \frac{\hat{\tenX} \btimes_1 \v_k \btimes_2\v_k }{\| \hat{\tenX} \btimes_1 \v_k \btimes_2 \v_k\|}.$
 \item $\v_k = E_{max}(\hat{\tenX} \times_3 \u_k).$
\end{enumerate}
\item CP Scaling: $d_k = \hat{\tenX} \times_1 \v_k \times_2 \v_k \times_3 \u_k \times_4 \w_k$. 
\item Projection: $\Pmat_{k} = \mathbf{I} - \V_{k}\V_{k}^{T} $.
\item Deflation: $\tilde{\tenX} = \tilde{\tenX} - d_{k}
  \v_{k} \circ \v_{k} \circ \u_{k} $
\end{enumerate}
\end{enumerate}
\end{algorithm}

When applied to tensor brain networks, our new semi-symmetric tensor
decomposition results in a method for TN-PCA.
Specifically, each $\u_{k}$ denotes the {\it subject-mode} and each
$\v_k \v_k^{T}$ denotes the rank-one {\it network mode}.  
The subject-modes give the low-dimensional vector embeddings of the
brain network for each subject; we use these to associate structural
connectomes with traits.  We call the weighted sum of network modes,
$\sum_{k=1}^{K} d_k \v_k \v_k^{T}$, the principal brain network, which
gives a one network summary that captures the most variation in the
structural connectomes across all subjects.  

\subsection{Relating Connectomes to Traits}

The TN-PCA $\tenX = \sum_{k=1}^{K} d_{k} \v_{k} \circ \v_{k} \circ \u_{K}$ embeds the weighted networks from $N$ subjects into vectors $\U_K = [\u_1,...,\u_k]$, where each row represents a low-dimensional representation of  a weighted network in $\real^K$. Since we have restricted $\{\v_k\}$ to be orthogonal to each other, we have $(\v_k \circ \v_k) * (\v_{k'} \circ \v_{k'}) = {\bf 0}$ for $k \neq k'$. Therefore, another way to view TN-PCA is that $\{\v_k \circ \v_k\}$ are basis networks and $\u_k (i)$ for $i = 1,...,N$ are the corresponding normalized coefficients (for $i$th subject) and the scale is absorbed by $d_k$.  

To relate the networks to various traits, we rely on the low-dimensional representations $\U_K$.  There are many advantages of using the vector $\U_K(i,:)$ to represent the $i$th subject's brain connectivity, e.g.,  (i) we recover the network from $\U_k(i,:)$ through $\sum_{k=1}^{K} d_{k}* \U_{k}(i,k)* \v_{k} \circ \v_{k}$; (ii) the low-dimensional vector representations bring us flexibility to utilize various existing statistical tools to study the relationship between the brain networks and human traits; and (iii) the TN-PCA is flexible and can be easily extended to deal with high-order tensors, e.g. a four-mode tensor, by simply include another vector in the outer product. 

{\bf Hypothesis testing of distribution difference}:  For each trait, we sort the 1065 HCP subjects according to their scores, and extract two groups of subjects: 100 subjects with the highest trait scores and 100 subjects with the lowest scores. For discrete traits, it is sometimes not possible to identify exactly 100 subjects; in such cases, we randomly select subjects on the boundary as needed.  We compare the embedded vectors of networks (rows of $\U_K$) from the two groups and test the null hypothesis that the two samples are from the same distribution against the alternative that they are from different distributions. We use the Maximum Mean Discrepancy (MMD) \citep{gretton2012kernel} to perform hypothesis tests. FDA is controlled using \cite{benjamini1995controlling}. 

{\bf Relating brain connectomes with traits}:  We are interested in studying relationships between traits and brain connectomes for all HCP subjects. In particular, we are interested in two types of results.   First, we want to see if brain structural connectomes can be used to predict various traits. Second, for those traits that can be predicted by brain connectomes, we would like to identify how the connectome changes with trait values by flagging the subset of connections with the largest differences.

Let ${\bf Y} = [y_1,...,y_N]^T $ be a $N \times 1$ vector, which contains one type of trait scores for each of the $N$ subjects, and let $\U_K$ be an $N \times K$ matrix containing the embedded networks in $\real ^K$ for $N$ subjects.  Our first analysis focuses on predicting $y_i$ using  brain PC scores and demographic covariates such as age and gender. Subjects in the HCP are randomly allocated into a training data set (66\% of the subjects), a validation data set (17\% of the subjects) and a test data set (17\% of the subjects). We then train various machine learning methods (e.g., simple linear/logistic regression, random forests, support vector machines and XGboost) to predict the trait scores. The prediction accuracy is evaluated using the root-mean-square error (RMSE) for both continuous and ordinal traits and  the classification accuracy for categorical traits (the trait type is identified manually, see the Supplement II). To evaluate whether brain connectomes are important in predicting traits, we compare with a reduced model, where only demographic covariates are used as predictors. The model containing brain connectomes is referred as the full model, and the model without brain connectomes is referred to as the baseline model. 

For each trait, we define a measure $\rho$ to evaluate the importance of brain connectomes. For trait $p$, let $\psi_p^f$ denote the RMSE or (1- classification accuracy) of the full model, and $\psi_p^b$ for the baseline model.  The  measure $\rho$ for trait $p$ is calculated  as $(\psi_p^b - \psi_p^f)/\psi_p^b$. The best model (including the tuning parameters in each machine learning method) is selected based on the validation data set, and then is applied to the test data set for performance evaluation.  We then evaluate $\rho$  for different combinations of networks and traits. 

The next question that we are interested in is, for each trait, how the connectome varies across levels of the trait? Depending on a trait's type, we apply different methods to identify the subset of networks or edges. For continuous traits we use canonical correlation analysis (CCA) \citep{hotelling1936relations} and for categorical traits we use linear discriminant analysis (LDA) \citep{fisher1936use}.
For a continuous trait,  the problem of finding a subset of edges that are highly correlated with the trait using CCA is equivalent to finding a direction $\bf w$ in  $\real^K$ (in the network embedding space) such that the correlation between the trait scores $\{y_i\}$ and the projection scores $\{ {u}_{proj}(i) = \left<\U_K(i,:),{\bf w} \right> \}$ are maximized, i.e. 
\begin{equation}
\label{equ:pls}
 \mbox{argmax}_{{\bf w} \in \real^K} \mbox{COV} (y,u_{proj}) = \mbox{argmax}_{{\bf w} \in \real^K}  \frac{1}{N} {\bf w}^T \U_K^T {\bf Y} ~~~~~~ \mbox{ s.t. } {\bf w}^T * {\bf w} = 1.
 \end{equation}
If we assume that both ${\bf Y}$ and rows of $\U_K$ are centered, the unit vector ${\bf w}$ obtained in (\ref{equ:pls}) describes changes of networks in the embedding space with increases of the trait score. To map this direction ${\bf w}$ back on to the brain network for interpretability, we let $\Delta_{net} = s\sum_{k=1}^K d_k {\bf w}(k) {\bf v}_k \circ {\bf v}_k$,
where $d_k$ and ${\bf v}_k$ come from the TN-PCA analysis, and $s$ is a scaling parameter (which will be explained in the next paragraph).  Confounding influence of age and gender can be regressed out from $y$ before fitting the model in (\ref{equ:pls}). For a categorical trait, e.g., $y_i \in \{0,1\}$, we use LDA to identify edge changes from low to high scores. The idea is to find a $\bf w$ in  $\real^K$  that best separates the two classes of networks. Let ${\bf \mu}_0$ and ${\bf \mu}_1$ be the means, and $\Sigma_0$ and $\Sigma_1$ be the covariances of the embedded networks.  The separation of the two groups is defined in the embedding space in the following way:
\begin{equation}
\label{equ:lda}
S = \frac{\sigma^2_{\text{between}}}{\sigma^2_{\text{within}}} = \frac{({\bf w}^T {\bf \mu}_0 - {\bf w}^T {\bf \mu}_1)^2}{{\bf w}^T (\Sigma_0 + \Sigma_1) {\bf w} }.
 \end{equation}
It is clear that $S$ achieves the maximum when ${\bf w} \propto (\Sigma_0 + \Sigma_1) ^{-1} (\mu_1 - \mu_0)$.  

In both CCA and LDA methods, we only identify a unit vector ${\bf w}$, reflecting the global trend in network changes with increasing traits, and there is no scale information included in ${\bf w}$. This makes the comparison of network changes across different traits challenging.   Hence, we want to define a proper scale $s$ for each trait such that $\Delta_{{net}} $ reflects the magnitude of network changes with increases in that particular trait. Since in this paper the  ${\bf w}$'s are always inferred from two groups of subjects (with high and low trait scores even for continuous traits), we dichotomize the subjects and use their mean differences in the network embedding space to define $s$. To be more specific, for a selected trait, letting $\bar{\bf u}_0 , \bar{\bf u}_1 \in \real^K$ be the mean embedding vectors for subjects with low and high trait scores respectively, we define $s = \|\bar{\bf u}_0 - \bar{\bf u}_1 \| $. 

\section{Results}
\subsection{Comparative studies of TN-PCA model}
In this section, we compare the proposed TN-PCA via CP
decomposition with other tensor decompositions, namely the Higher-Order
SVD (HOSVD) and Higher-Order Orthogonal Iteration (HOOI)
\citep{tucker_1966,de_lathauwer_2000,kolda_tensor_review}.

First, we study how well the HOOI, HOSVD and our new CP algorithm for
semi-symmetric tensor decompositions perform in simulations.  
We simulate data from the following semi-symmetric tensor model:
$\tenX = \tenD \times_{1} 
\V \times_{2} \V \times_{3} \U + \mathbf{\mathcal{E}}$, where $\tenX$ is a $P\times P \times N$ semi-symmetric tensor and 
$\mathbf{\mathcal{E}}_{\cdot, \cdot,l} \overset{iid}{\sim}
\textrm{Wishart}(\mathbf{I},P)$ to ensure the noise follows a
semi-symmetric structure.  Also, we take $\tenD$ to be a diagonal
tensor to most fairly compare the CP and Tucker models, letting
$\tenD_{k,k,k} = (2 - 0.1k)*\sqrt{PM}$.  For the tensor factors, $\V$
is randomly sampled from the Stiefel manifold; $\U$ is either
generated from the Stiefel manifold
or generated as non-orthogonal iid norm-one Gaussian vectors to evaluate both the orthogonal and non-orthogonal
$\U$ cases.  Each of our scenarios is studied for a set of tensors with size $P=100$, $M=500$ for various ranks $K$,
and under various  signal-to-noise (SNR) levels, where the SNR is defined as ${\| \tenD
  \times_{1} \V \times _{2} \V \times_{3} \U \|_2 }/{ \| {\boldsymbol{\mathcal{E}}} \|_2}$.  We evaluate each tensor
decomposition algorithm according to two metrics: the relative
difference between the true $\tenD$ and estimated $\hat{\tenD}$, $\|\tenD - \hat{\tenD} \|_2/\|\tenD\|_2$, and the cumulative proportion
of variation explained by the first $K$ components.  We define this as
follows: Let ${\bf V}_k = [\v_1,...,\v_k]$ and ${\bf P}^{\bf V}_k = {\bf
   V}_k ({\bf V}_k^T{\bf V}_k)^{-1}{\bf V}_k^T$; define $\U_k$ and
$\Pmat^{\U_k}$ analogously.  The cumulative proportion of variation explained
 by the first $k$ high-order components is  
$ \| \tenX \times_1 {\bf P}_k^{\bf V} \times_2 {\bf P}_k^{\bf V}
 \times_3 {\bf P}_k^{\bf U}\|^2 / \| \tenX\|^2$ \citep{allen2012sparse}. 
 All simulations were repeated ten times with the average results reported.

\begin{figure}
\small
\begin{center}
\begin{tabular}{|cc || cc|}
\hline
\multicolumn{2}{|c||}{Orthogonal $\U$}&\multicolumn{2}{c|}{ Non-orthogonal $\U$} \\
\hline
$\|\tenD - \hat{\tenD} \|_2/\|\tenD\|_2$ & Variance explained &  $\|\tenD - \hat{\tenD} \|_2/\|\tenD\|_2$ & Variance explained \\
\hline
\includegraphics[height=1.1in]{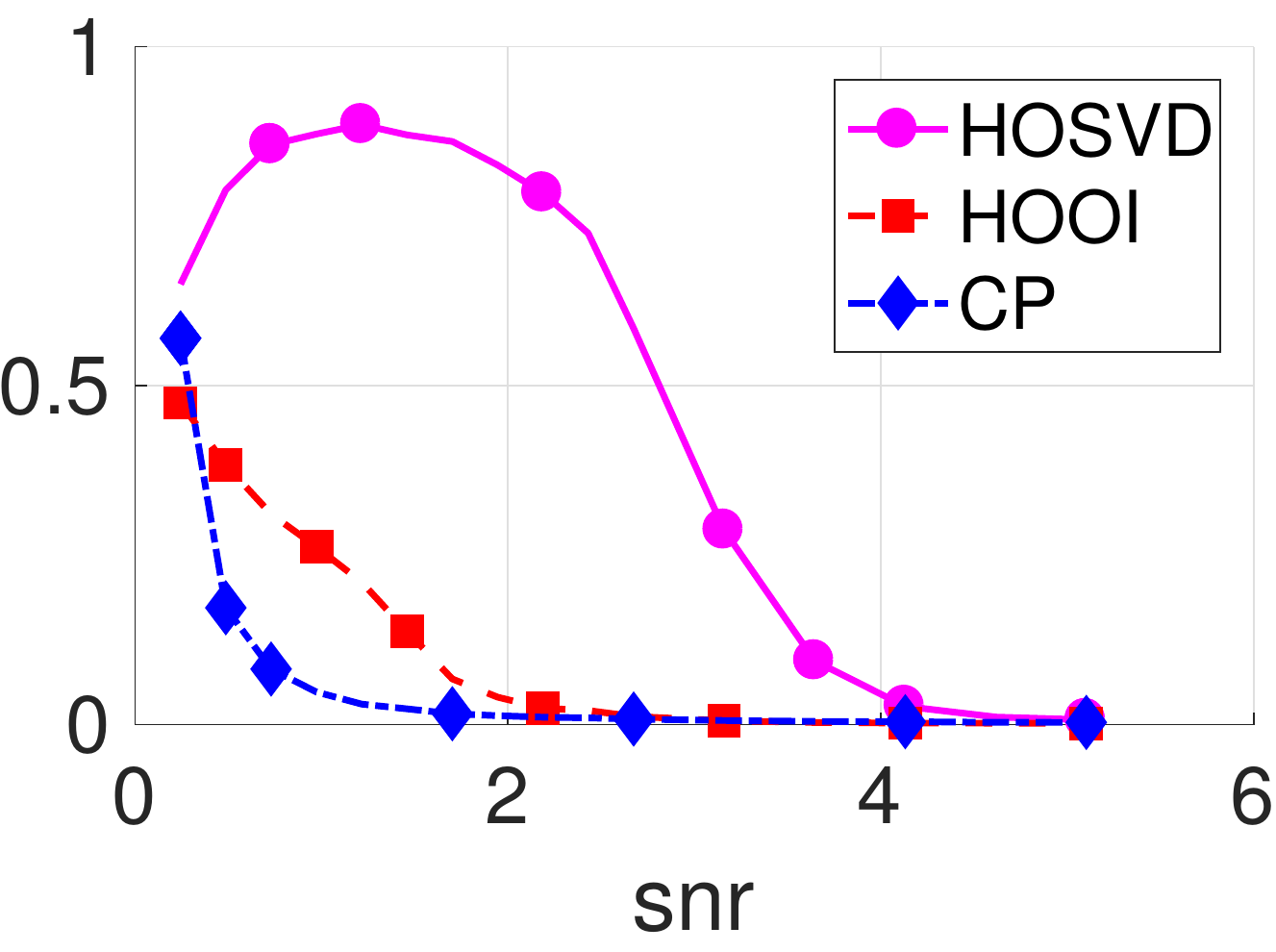} &
\includegraphics[height=1.1in]{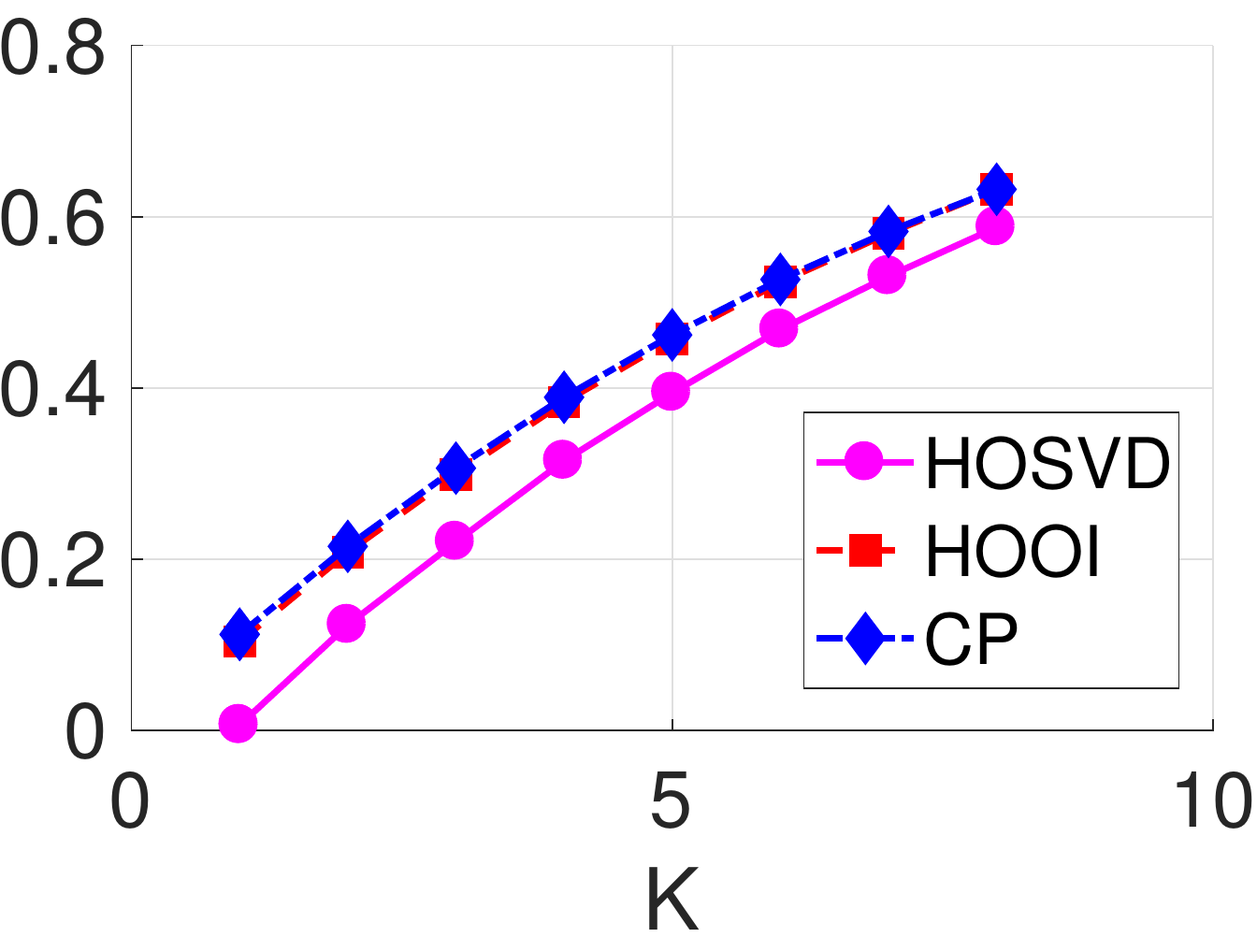} &
\includegraphics[height=1.1in]{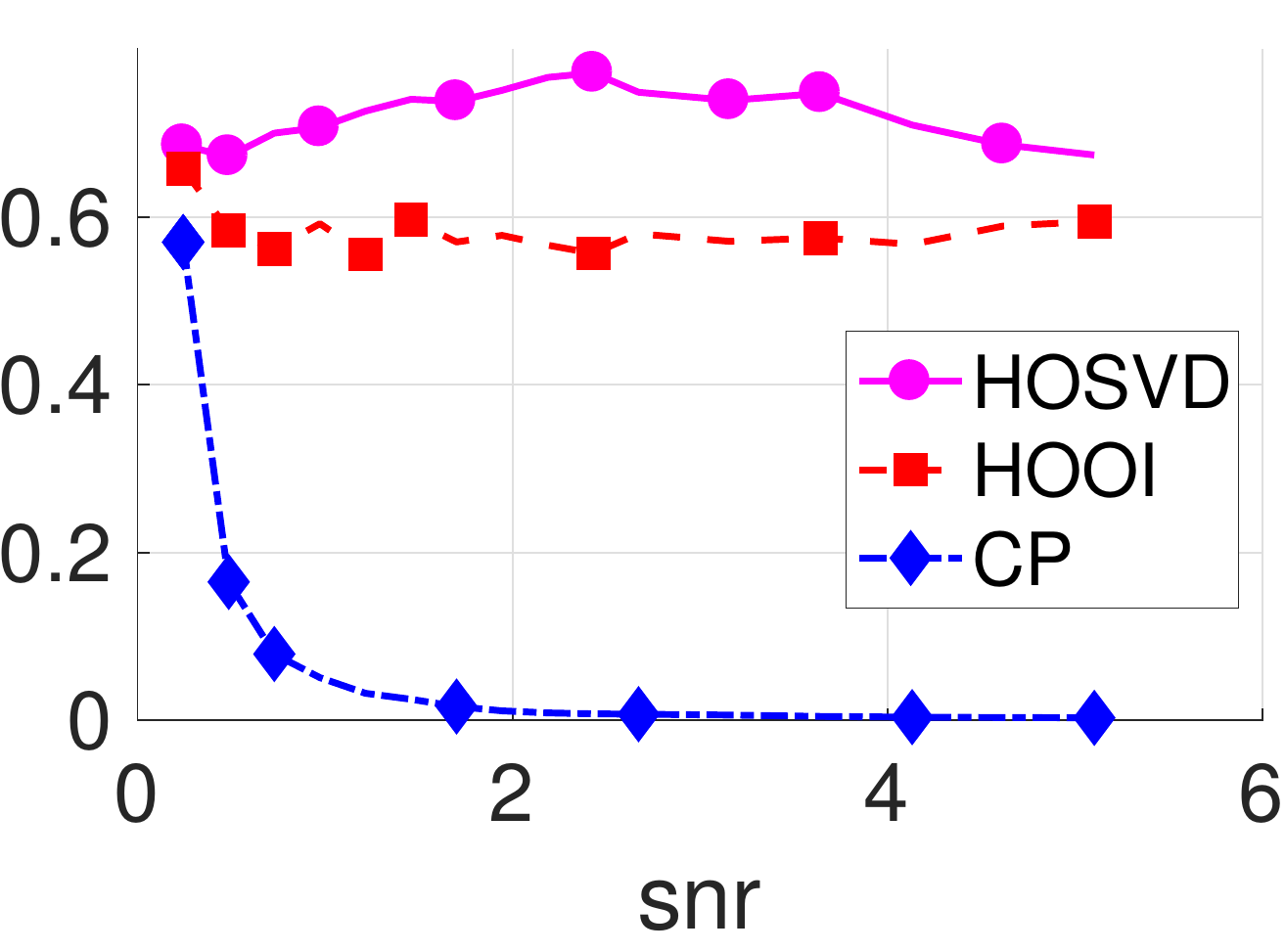} &
\includegraphics[height=1.1in]{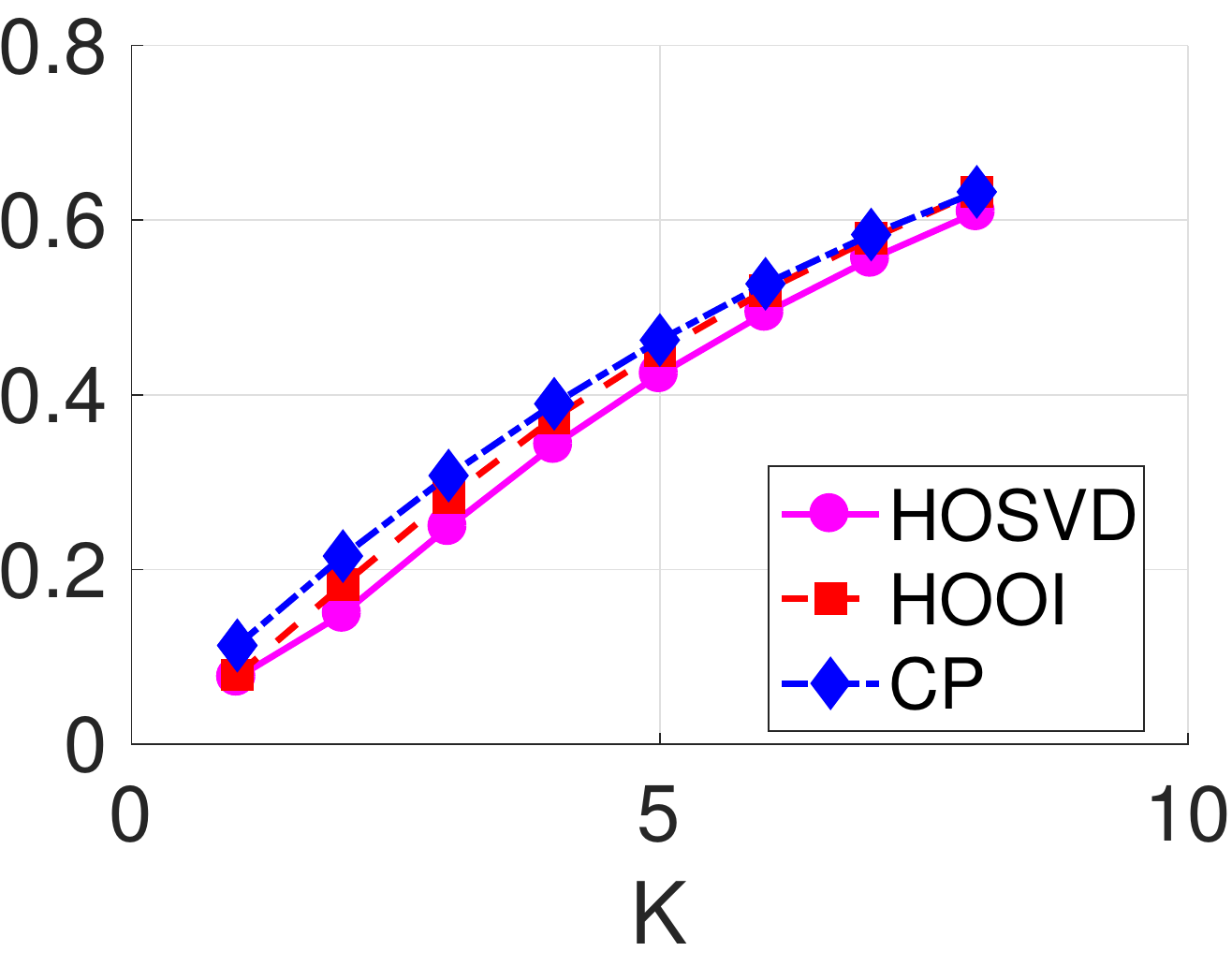} \\
\hline
\end{tabular}
\caption{Simulation results in terms of structural recovery (left columns) and
  variance explained (right columns) for the proposed TN-PCA via a new semi-symmetric CP
  model compared to the existing HOSVD and HOOI methods.} 
\label{fig:pcs1}
\end{center}
\end{figure}

The results presented in Figure 1 show that our new CP model for
semi-symmetric tensors compares favorably to the algorithms based on
the Tucker model, the HOSVD and HOOI.  This is especially true in terms
of structural recovery (left columns), where our model is able to
recover the structure at lower SNRs and when the $\U$ factors are not
orthogonal.  In terms of variance explained (right columns),
our CP model shows marginal improvements over the Tucker model;
the added flexibility of the non-diagonal Tucker core may help  
explain more variance in the data even if the lower-dimensional
structural recovery is poor. In a real data analysis presented in Supplement I, we also demonstrate the flexibility of the proposed TN-PCA. HOOI requires carefully tuning $K$ to avoid lack-of-model fit issues, but our new semi-symmetric CP model does not suffer
from this problem (we can simply choose $K$ that
explains most of the variance in the data without worrying
about model mis-specification).

\subsection{Exploratory analysis: connectome visualization and classification}

The TN-PCA approach approximates the brain tensor network using $K$ components, with the components ordered to have decreasing impact.  Individuals are assigned a brain connectome PC score for each of the $K$ components, measuring the extent to which their brain tensor network expresses that particular tensor network (TN) component.  In this section, we demonstrate how to use these PC scores for visualization of brain connectomes in large cohort studies and statistical analysis of connectomes, such as classification.

\begin{figure}
\begin{center}
\begin{tabular}{c}
\includegraphics[height=2.4in]{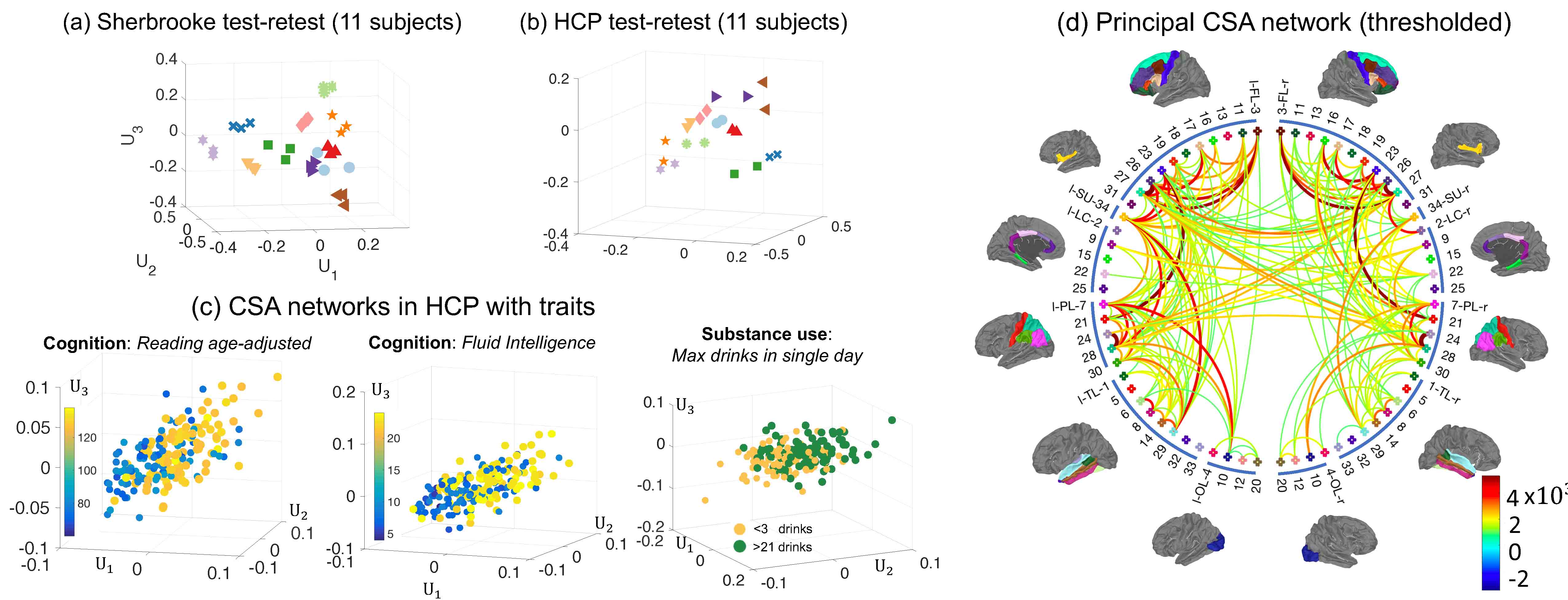} 
\end{tabular}
\caption{Illustration of connectome visualization using Tensor Network (TN-) Principal Components Analysis (PCA). In (a) and (b), we display brain PC scores for CSA networks from the HCP and Sherbrooke test-retest data set. Each unique combination of color and marker type represents multiple scans from the same subject (3 scans per subject in Sherbrooke and  2 scans per subject in HCP).  In (c), we show PC scores with traits.  For each cognition trait, we selected $100$ subjects with low trait scores and $100$ subjects with high scores. For the substance use trait (alcohol use), we selected subjects with $<3$ drinks and subjects with $>21$ drinks.   In (d), we display the principal CSA brain network calculated using all 1065 subjects from the HCP data set. For display purposes, we thresholded the dense principal brain network to keep only the 200 most connected edges. } 
\label{fig:embedding}
\end{center}
\end{figure}

\begin{figure}
\begin{center}
\begin{tabular}{c}
\includegraphics[height=2.4in]{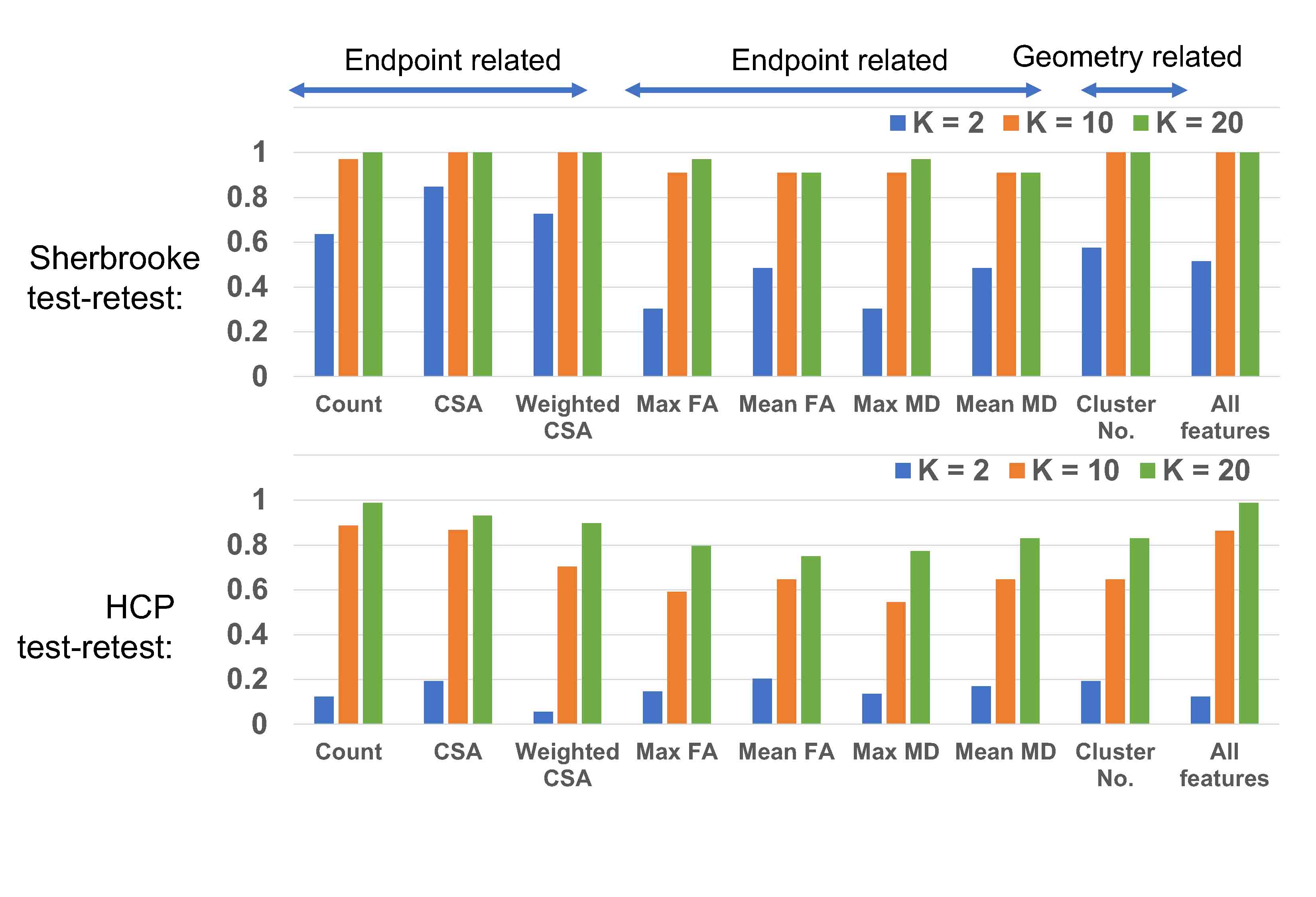} 
\end{tabular}
\caption{Classification results on the Sherbrooke and HCP test retest data sets. The results are based on nearest neighbor classification. }
\label{fig:classification}
\end{center}
\end{figure}

 In the following study, we use  the CSA network as an example. Figure \ref{fig:embedding} (a) and (b) shows brain PC scores for each of CSA networks in the Sherbrooke and HCP test-retest data set using $K=3$.  Each combination of color and marker type represents multiple scans from the same subject (3 scans per subject in Sherbrooke and  2 scans per subject in HCP). Even using only $K=3$ components (for data visualization), the CSA brain networks display a clear clustering pattern, suggesting that not only are the extracted connectomes from the repeated scans reproducible but also that we can distinguish between different subjects based on only three components. To formally assess this, we applied nearest neighbor clustering to the PC scores for different types of weighted networks separately (3-way tensor decomposition) and jointly (4-way tensor decomposition) under different $K$. Figure \ref{fig:classification} shows the results (some non-discriminative features are excluded based on the results in supplementary Figure 1).  
 
 We have the following interesting observations: 1) with a moderate $K$, e.g, $K=10$, we can almost perfectly cluster the repeated scans of all subjects based on the discriminative features such as count, CSA, and cluster number; 2) among the three classes of features, the endpoint-related features have more discriminative power than the other two sets; and 3) we can obtain very good classification rates by jointly using all features.  Although the HCP data have better resolution, it seems that the classification results of HCP are not better than for Sherbrooke. There are several reasons. First, there are more subjects (44 vs. 11) and less scans (2 vs. 3) per subject in the HCP than Sherbrooke. More subjects with less scans per subject increases the variation in the data, and thus requires more PC coefficients to well represent each network and it also makes the classification difficult. Second, scan intervals in the HCP (about 60\% of them are $>$ 5 months) are much larger than Sherbrooke (1-2 weeks). Differences between scans of a subject in the HCP test retest is naturally larger than Sherbrooke.

TN-PCA allows one to visualize relationships between the structural connectome and various traits in the HCP data. Based on CSA networks, Figure \ref{fig:embedding} (c) displays the first three brain PC scores along with three selected traits (two cognition, one substance use). For the two cognition traits (oral reading test score and fluid intelligence), we selected $100$ subjects with low trait scores and $100$ subjects with high scores and plotted their brain PC scores. For the substance use trait (max drinks in a single day), we plot subjects with low ($<3$ alcohol drinks) and high values ($>21$ alcohol drinks). We can clearly observe separation between different groups of subjects in these plots,  indicating that brain connection patterns are different for these two groups (measured by the CSA feature; we have similar findings on some other features, e.g. the fiber count). 

Principal brain networks can also be obtained as a byproduct of
TN-PCA.  We define the rank $K$ principal brain network to be the network
given by the sum of the first $K$ rank-one tensor network components from
TN-PCA.  Similar to
examining patterns amongst features by exploring the PC loadings, the
principal brain network exhibits major patterns of structural
connectivity that explain most of the variation across the population.
Figure \ref{fig:embedding} (d) displays the thresholded principal
brain network derived from the CSA networks of 1065 subjects from the
HCP data set; here, we take $K=30$ and threshold the edges so that
the 200 most connected pairs of brain regions are displayed.  
The principal brain network gives a visual summary of all the
structural connectomes in the HCP population.  Compared to the mean
network for the HCP population (shown in
Supplementary Figure 2), our principal brain network
yields additional insights into which structural connections tend to
have major differences across subjects.  Specifically, we see that
there is large variation in the strength of connections within
hemispheres as well as a few strong inter-hemispheric connections 
that vary across subjects.   As we will investigate in the next sections, these variations in brain connections across subjects may be related to differences in traits.

\subsection{Statistical inference: relating connectome to traits}
In this section, we study distribution differences of brain connectomes across different groups, evaluate the prediction power of brain connectomes and infer how the connectome varies across levels of the trait. 

\begin{figure}
\begin{center}
\begin{tabular}{c}
\includegraphics[height=2.8in]{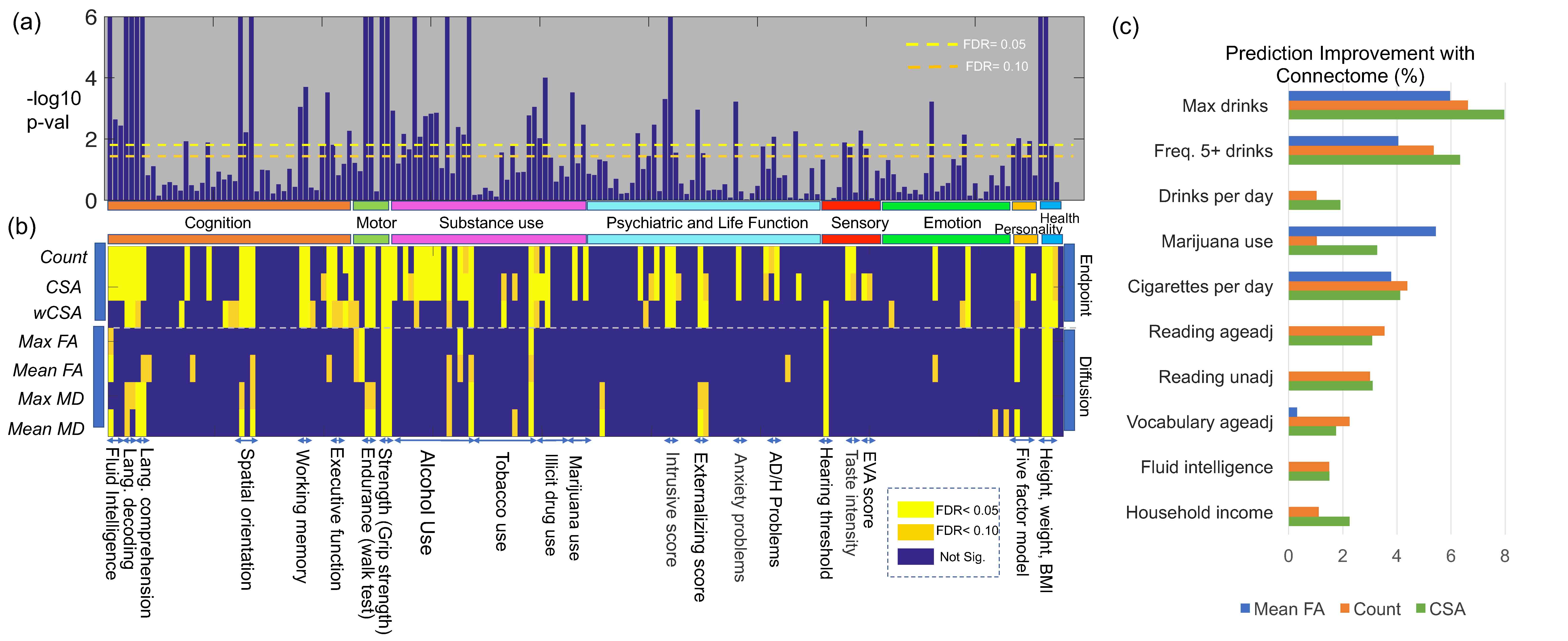} 
\end{tabular}
\caption{Relation between structural brain connectomes and various traits. Panels (a) and (b) show hypothesis testing results for $175$ traits and $7$ different weighted networks. In (a), we show p-values ( $-\text{log}_{10}$ scale) of the CSA weighted networks with different traits. Two different FDR thresholds are used (different colored dash lines). In (b), we present hypothesis testing results for different combinations of traits and networks.  Each row shows a type of weighted network and each column shows a particular trait. The significance is displayed based on different FDR values. Panel (c) shows the top 10 traits in terms of their predictability based on brain connectomes adjusting for age and gender as covariates.  The prediction improvement ratio $\rho_p$ for the $p$th trait is shown.}
\label{fig:hyptest}
\end{center}
\end{figure}

{\bf Hypothesis testing of connectome distribution difference:}  We first assess whether there are significant differences in distribution of brain PC scores among subjects having a low versus high value for each trait.  Out of the 1065 HCP subjects, we identified groups of 100 having the highest and lowest values for each trait.  We used the Maximum Mean Discrepancy (MMD) test \citep{gretton2012kernel} to obtain $p$ values for differences in the brain connectomes across these two groups for each trait.  Results are shown in Figure \ref{fig:hyptest}. Figure \ref{fig:hyptest} (a)  shows $p$ values for the CSA weighted networks.  Different thresholds for significance based on false discovery rate (FDR) control using \cite{benjamini1995controlling} are marked with different colored lines.  Corresponding results for 7 types of weighted networks are shown in  Figure \ref{fig:hyptest} (b). 

Based on these results, many traits are significantly related to brain structural connectomes. Traits in the same domain are placed next to each other in Figure \ref{fig:hyptest} (b).  The block patterns of significance add evidence that brain structural connectomes relate more broadly to these trait domains.
In the cognitive domain, structural connectomes are related to  fluid intelligence,  language decoding and comprehension, working memory, and some executive functions. In the motor domain, connectomes are related to endurance and strength. In the substance use domain, connectomes are related to alcohol consumption, tobacco, illicit drug and marijuana use. In the sensory domain, connectomes are related to hearing and taste. In the emotion domain, connectomes are related to negative emotions, such as anger and anxiety. In the health domain, connectomes are related to height, weight and BMI. Again, we see that endpoint-related features are more discriminative according to the hypothesis testing results; more traits are significant adjusting for false discoveries using features such as count, CSA and weighted CSA than when using diffusion-related features.


{\bf Prediction of traits using connectomes:}  We are first interested in comparing our global structural connectome with other types of connectomes, such as local structural connectome \citep{powell2018local,yeh2016connectometry} and functional connectome, in teams of predicting human traits. To directly compare with the local structural connectome, we utilize a subset of traits that have been used in \cite{powell2018local} in a similar setting: a linear regression model is applied within a cross-validation paradigm (5-fold cross-validation) with $K = 60$. For functional connectome, we utilize the Desikan-Killiany atlas and the resting state fMRI data to calculate a $68\times 68$ correlation matrix for each subject and then the same procedures (TN-PCA and linear regression) are applied to evaluate the prediction power.  The results are shown in Table \ref{tab:compfc}, where column 4 shows the correlation between the predicted and the observed traits and column 5 shows the p-value of the correlation being zero for the count structural connectome. We can see that our structural connectome has better prediction ability than the local structural connectome. When compared with functional connectome, structural connectome is better for most traits, but a few traits can be predicted better with functional connectome. 

\begin{table}ett
\caption{Comparison of trait prediction results based on different brain connectomes.}
\label{tab:compfc}
\footnotesize
\begin{tabular}{p{4cm }|p{3cm} | p{3cm} | p{3cm} c }
\hline
 Model Response & {Local structural connectome} (correlation)  & {Functional connectome}(correlation) &   {Global structural connectome} (correlation) & P-value \\
\hline
Total household income & -0.0029 & 0.1279 &  0.1342& 0.0510 \\
Years of education completed  & 0.0729 & 0.0619 &0.1536 & 0.0257$^*$ \\
 Picture Sequence  Memory Test & 0.0977& 0.1066&0.0671 & 0.3306 \\
Card Sort Test &-0.0299 & 0.0591&0.0522 &  0.4493 \\
Inhibitory Control and Attention Test &-0.0001 & 0.1120&0.0931& 0.1760\\
Penn Progressive Matrices:
Number of correct responses  & 0.0849 &0.2400 & 0.2411& 0.0004$^{**}$ \\
Penn Progressive Matrices:
Total skipped items &0.0733& 0.1976&0.2106& 0.0021$^{**}$ \\
Penn Progressive Matrices:
Median reaction time for
correct responses &0.0086 & 0.1125 &0.1218  & 0.0769\\
Oral Reading Recognition Test &0.0008  &0.2393 &0.3020& $<$0.0001$^{**}$\\
Picture Vocabulary Test & 0.0481 & 0.2091&0.2622  &  $<$0.0001$^{**}$\\
Processing Speed Test  &  -0.0569 & 0.0891&0.0552 & 0.423\\
Delay Discounting: Area under the curve for discounting of $200$  &0.0275  & 0.0808&0.1483  & 0.0309$^*$\\
Delay Discounting: Area under the curve for discounting of $4000$  & 0.0802 &0.2268 &0.0911  & 0.1864\\
Line Orientation: Total number correct &0.0951  & 0.1254&0.2536 &0.0002$^{**}$\\
Line Orientation: Median reaction time divided by expected number of clicks for correct  &-0.0572 &0.0243 &0.0974  & 0.1578\\
Line Orientation: Total positions off for all trials  & 0.0014& 0.2191&0.2248  &0.0010$^{**}$\\
Memory Test: Total number of correct responses  & 0.0474& 0.0273&0.1034 &0.1335\\
Memory Test: Median reaction time for correct responses  &-0.0391 & 0.0473&0.0404 & 0.5589\\
List Sorting Working Memory Test  &0.0793 &0.1121 &0.0939  &0.1723\\
Body mass index &0.2736 & 0.2770& 0.1933 & 0.0047$^{**}$ \\
Sleep Quality Index &  -0.0314 & 0.0747&0.0455  & 0.5092\\
\hline 
\end{tabular}
\raggedright Correlations in the first column are copied from \cite{powell2018local}, whose experiment setting is similar to us but with less subjects. In the last column, * indicates significance based on $\alpha = 0.05$ and ** indicates significance after FDR based on $\alpha = 0.05$. 
\end{table}

To better study the relationship between connectomes and traits and to account for some potential confounding effects, we consider a more comprehensive strategy:  a baseline model that only uses demographic variables of age and gender as predictors is compared with a full model that also includes brain connectome PC scores. In this scenario, the $1065$ subjects in the HCP data set are randomly divided into three groups: a training group containing $66\%$ of the subjects, a validation group containing $17\%$, and a test group containing $17\%$. We trained various machine learning algorithms for prediction using the training data set, with the number of components $K$ treated as a tuning parameter.  The prediction  improvement of the full model over the baseline model is measured using the relative ratio $\rho$.  For each trait, the best model (with the highest $\rho$) is selected based on the validation data set. {Panel (a) of supplementary Figure 3 presents the results} ($\rho$'s for different traits and features) for the validation and test data sets based on the average of $50$ runs. According to the values of $\rho$ in the
test data set, we selected the 10 traits yielding the largest
predictive improvements based on connectomes. The $\rho$'s for these
10 traits are displayed in Figure \ref{fig:hyptest} panel (c). The 10
traits come from two domains: substance use (5 traits) and cognition
(5 traits). 

Among the five traits of substance use, three of them are
related to alcohol use, one to cigarette use and the last
one to marijuana use. A close inspection of the two
alcohol use traits was performed and the results are presented in the
{Supplementary Figure 3 panels (c) and (d)}. Consider the trait that
measures lifetime max drinks in a single day as one example; it is an
ordinal variable ranging from 1 to 7, with 1 corresponding to less
than 3 drinks and 7 to more than 21 drinks. For subjects with reported
values from 1 to 7, our model based on the brain connectome PC scores
predicted mean values (on the test data set) for each group of $1.97,
1.83, 2.27, 2.67, 2.91, 3.13, 4.04$, respectively, showing a clear
increasing pattern. In another example presented in Figure 4, we
extract subjects with value 1 (light drinkers, totalling 191) and 7
(binge drinkers, totalling 93) and use LDA to perform binary
classification.   Based on the brain connectome PC scores alone, we
obtain a classification accuracy of $80.99\%$.  These results 
suggest that using only the brain connectomes, we can distinguish
with surprising accuracy between individuals with low and binge alcohol
consumption.  

Of the five leading cognitive traits, three are related to language and vocabulary decoding ability, one to fluid intelligence, and the other to household income (we loosely classify the household income into the cognition category).  {A close inspection of the language decoding trait is presented in the Supplementary Figure 3 panel (b).} The language decoding trait score (after age adjustment) can be predicted $\sim 4\%$ better ($p<0.0002$) under a random forest model with the additional CSA brain PC scores (the random forest model is selected based on validation data). On the test data set, the correlation between the predicted trait and the subject self-reported trait is $r=0.27$ (based on the average of ten runs). If we restrict the analysis to the 200 subjects with the highest and lowest traits (plotted in the first column of panel (c) in Figure \ref{fig:embedding}), the correlation increases to $r=0.45$.   Similar results are observed for the traits of fluid intelligence and vocabulary decoding. Given that these trait scores are only a rough measure of a person's cognitive function, the results are very promising, indicating that brain structural connectomes can partially account differences in cognitive abilities.

{\bf Interpreting relationships between traits and connectomes:}
Having established that a particular trait is significantly associated and can be predicted with the brain connectome, it is important to infer how the connectome varies across levels of the trait.  For example, is the association specific to certain sub-networks and in what direction is the association?  We start by studying how brain connectome PC scores vary with the trait, and then map these changes back onto the network using the method presented in the Methods section. The following analyses are based on $K = 30$ (results are robust for $K\sim20-60$).  We selected three representative traits: language reading age adjusted score, lifetime max drinks consumed in a single day and the use of marijuana. The language reading score is a continuous variable, and the other two are categorical. We study how the CSA network changes with increases in these traits.

\begin{figure}
\begin{center}
\begin{tabular}{c}
\includegraphics[height=2.5 in]{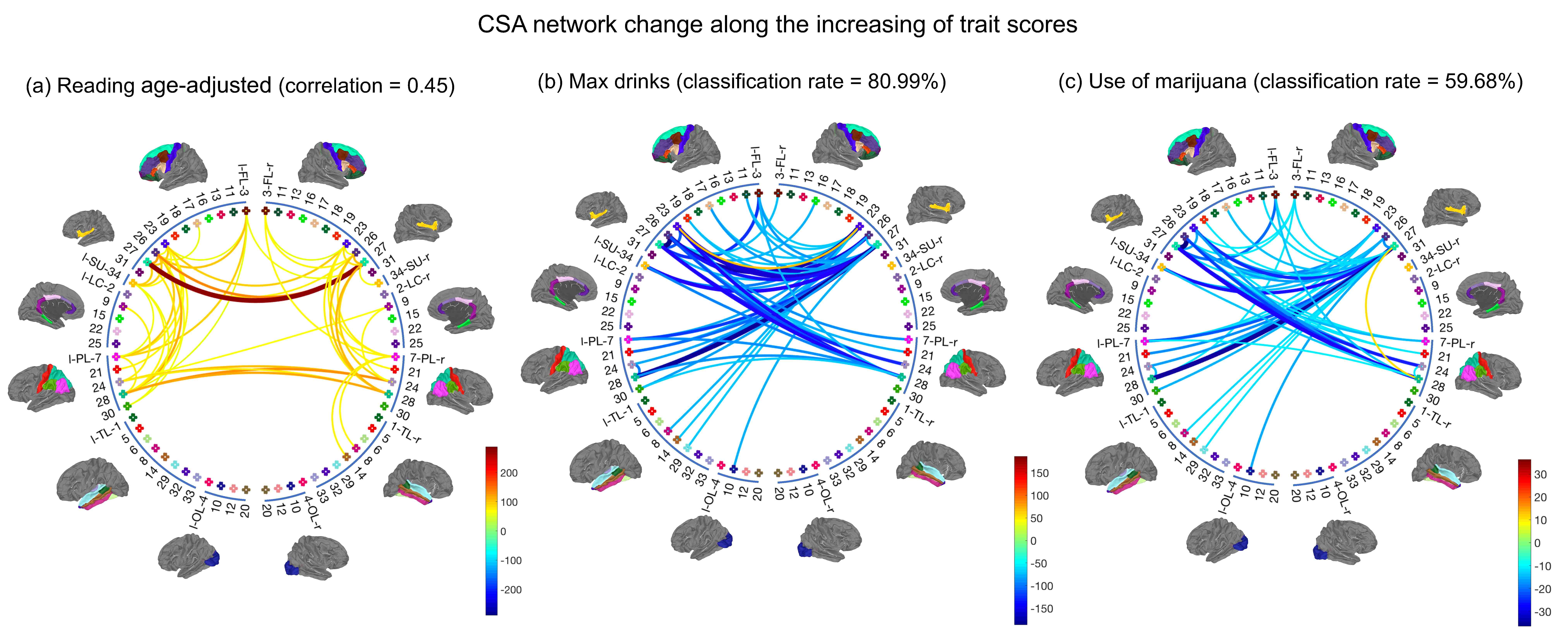} 
\end{tabular}
\caption{Top 50 pairs of brain regions in terms of their changes in CSA connectivity with increasing traits. (a) results for increasing language reading score. (b) results for max drinks; (c) results for marijuana use. 
We also display the correlation and binary classification rate obtained from CCA and LDA, providing measures of differences in brain connectivity networks between subjects with different trait scores.} 
\label{fig:plsresult}
\end{center}
\end{figure}

 For the language reading trait, the plot of the top 3 PC scores for the 100 subjects with the highest scores and 100 with the lowest scores is shown in panel (c) of Figure \ref{fig:embedding}. 
Panel (a) in Figure \ref{fig:plsresult} shows the network change ($\Delta_{net}$) with the trait; we show only the 50 edges that change most.  There are four major white matter connections that increase significantly with reading ability. They connect the left and right frontal lobes (FL) and parietal lobes (PL): (rFL, lFL), (rFL, rPL), (lFL, lPL), (rPL, lPL).   On closer inspection, the strongest connections are among left and right brain nodes of $27$-superior frontal, $26$-middle frontal (BA 46), $28$-superior parietal and $24$-precuneus ( Supplement III has detailed information on each ROI).  These regions have shown strong associations with language ability in previous studies of Positron Emission Tomography (PET), functional MRI \citep{edwards2010spatiotemporal,price2012review} and cortical thickness \citep{porter2011associations}.    There were no edges that decreased significantly with  increasing reading ability. This result is robust to adjustment for age and gender.

{Lifetime max drinks is an ordinal variable ranging from 1 to 7, with 1 corresponding to less than 3 drinks and 7 more than 21 drinks.  The third column in Figure \ref{fig:embedding} (c) shows the top 3 brain PC scores for individuals with a max drink score of 1 (191 light drinkers) and 7 (93 binge drinkers).  The separation between light and binge drinkers suggests structural connectome differences between the two groups.  Panel (b) in Figure \ref{fig:plsresult} shows the result of using LDA to identify brain network changes between light and binge drinkers ($\Delta_{net}$).  As max drinks increases, inter-hemisphere connections (especially the connections between rFL, lFL, rPL, and lPL) decrease. Different from the language reading trait, we observe that $\Delta_{net}$ for the alcohol drinking are mostly associated with the frontal lobes (almost all connections are involved with either rFL and lFL ). Previous studies \citep{moselhy2001frontal}  have provided evidence of relationships between alcoholism and dysfunction and deficits in the frontal lobe. We repeated this analysis for marijuana use (583 used and 481 did not), and the result is shown in panel (c) of  Figure \ref{fig:plsresult}.  Similarly to alcohol, marijuana use is also associated with decreases in inter-hemisphere connections.   }

To assess how well subjects with different trait scores can be distinguished based on their brain connectomes, we calculated the correlation between traits and their predictive values for continuous traits and the classification rate for categorical traits, based on $\left< {\bf w}, {\U}_K(i,:) \right>$.  The results are presented in Figure \ref{fig:plsresult}. The correlation between the projected ${\U}_K(i,:)$  and the language reading score is $0.45$ for the 200 subjects, indicating a strong relation between reading ability and the subjects' brain networks. The classification rate is $80.99\%$ for binge versus light drinking. More specifically, the sensitivity (binge drinker is identified as binge drinker)  is 59.1\% and the specificity (light drinker is identified as light drinker) is 91.6\%. The classification rate for marijuana use or not is $59.68\%$ (sensitivity: 26.8\%, specificity: 86.8\%)  This rate is surprisingly high for alcohol, indicating sizable differences in these subjects' brain structural connectomes, while the rate for marijuana is only slightly better than chance.

\begin{figure}
\begin{center}
\begin{tabular}{c}
\includegraphics[height=2.5 in]{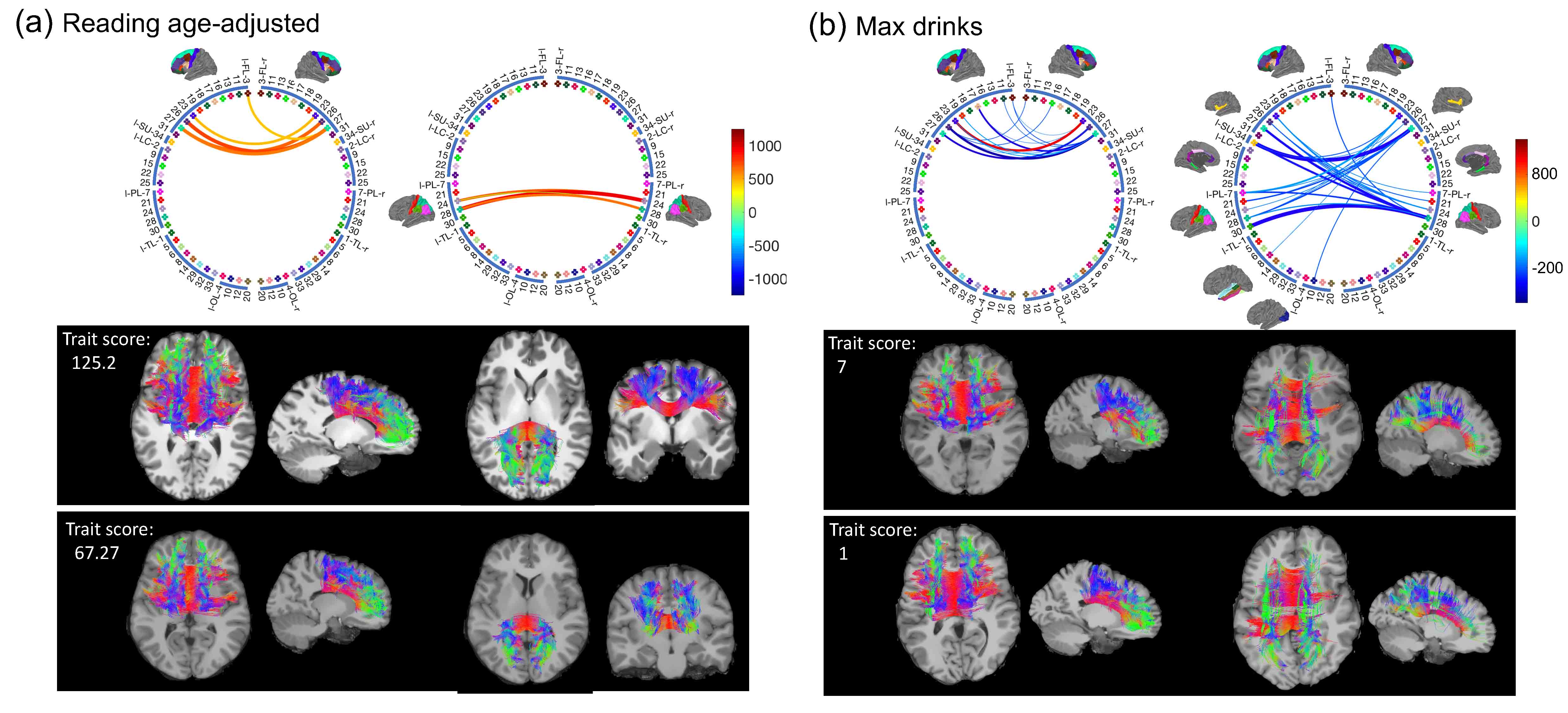} 
\end{tabular}
\caption{Network difference (based on CSA network) and the corresponding streamlines for selected subjects in the HCP data set.
Among the 50 pairs of brain regions identified in \ref{fig:plsresult}, we focus on cross-hemisphere connections.  Such connections are either (lFL, rFL) or (lPL, rPL).  In (a) we show the differences in these two 
sets between subjects with high (125.2) and low (67.27) reading scores. The streamlines corresponding to these connections are shown in the second row. Similar results are plotted in (b) for two selected subjects with light and binge drinking. } 
\label{fig:streamlineresult}
\end{center}
\end{figure}

From the networks presented in Figure \ref{fig:plsresult}, we extract their corresponding white matter tracts and display them in Figure \ref{fig:streamlineresult}. These white matter tracts are from two selected subjects with high and low trait scores. Among the connections shown in Figure \ref{fig:plsresult}, cross-hemisphere connections are particularly interesting.  These connections are roughly classified into two types: (lFL, rFL) and (lPL, rPL).  The first row of panel (a) in Figure \ref{fig:streamlineresult} plots differences in these two types of connections between subjects with high (125.2) and low (67.27) reading scores.  The streamlines corresponding to these connections are shown in the second row.  The subject with a high reading score has richer and thicker structural connections in both (lFL, rFL) and (lPL, rPL); this pattern is common for subjects with similarly high reading scores.
Figure \ref{fig:streamlineresult} panel (b) shows similar results for selected light and binge drinkers; in this case differences between the subjects are detectable but subtle.      { More detailed results for the two pairs of subjects are displayed in the Supplementary Figure 4 and 5.}

\section{Discussion}
Using state-of-the-art data science tools applied to data from the Human Connectome Project (HCP), we find that many different human traits are significantly associated with the brain structural connectome.  Overall, and consistent with results in previous studies of functional connectomes \citep{smith2015positive,finn2015functional}, positive attributes tend to have an overall positive association with structural connectomes, while negative attributes have a negative relationship.  Examples of positive traits include high language learning ability, fluid intelligence and motion ability; high levels of such variables tends to be indicative of stronger interconnections in the brain.  Examples of negative traits include a high level of alcohol intake or the use of marijuana; such variables tend to be indicative of weaker interconnections.  

Given inevitable errors in connectome reconstruction and in measuring human traits, such as alcohol intake, it is surprising how strong the statistical relationships are.  For example, we chose to highlight results for reading scores and alcohol intake as being particularly interesting.  For reading scores using our data science methods, the correlation between the measured reading score and our predicted value based on an individual's brain connectome was 0.45 (focusing on subjects with particularly low or high scores).  In addition, and even more remarkably, the classification accuracy in attempting to distinguish between a light drinker and an individual with a history of binge drinking based only on their brain connectome was surprisingly high.  ({See Supplementary Figure 5 for more results on alcohol}).  

Code for implementing the data processing pipeline, TN-PCA method of dimensionality reduction, and the corresponding statistical methods for prediction and interpretation are all freely available with documentation on GitHub (to be posted on acceptance of this paper). These methods should be highly useful in further studies for carefully studying relationships between brain connectomes and individual traits.  Careful follow-up studies are needed to better establish directions of causality.  For example, do individuals with less connected brains have more of a tendency for substance abuse or does substance abuse cause a decrease in connectivity?  The direction of this relationship has a fundamental impact on the clinical and public health implications of our results.  Also of critical importance is the plasticity of the connectome; for example, if a binge drinker modifies their drinking behavior does the brain gradually return to a normal connectivity pattern over time?  If an individual having a low reading score works hard to improve their score through coursework, tutoring and exercises, then does the brain connectivity also improve?  Does this intervention have a direct causal effect on the connectome?  

Given the increasing quality of data on structural connectomes, and in particular the sizable improvements in robustness and reproducibility, we are now at the point in which large, prospective studies can be conducted to answer some of the above important questions.  The tools developed and used in this article should be helpful in analyzing data from such studies.  On the methods side, it is important moving forward to continue to develop more informative and reproducible measures of connectivity between pairs of brain regions, and also to reduce sensitivity to the somewhat arbitrary number and choice of regions of interest.  An additional important direction is linking structural and functional connectivity together in one analysis, which can potentially be accomplished via a minor modification of the proposed TN-PCA approach. In addition, a more interpretable TN-PCA that allows sparsity will also be one of the future directions. 

\section*{Acknowledgements}

Data were provided in part by the Human Connectome Project, WU-Minn Consortium (Principal Investigators: David Van Essen and Kamil Ugurbil; 1U54MH091657). We also thank Maxime Descoteaux, Kevin Whittingstall and the Sherbrooke Molecular Imaging Center for the acquisition and sharing of the test-retest data. GA acknowledges support from NSF DMS 1554821 and NSF NeuroNex 1707400.

\section*{References}
\bibliography{paperdti,tensors,papertensor}

\end{document}